\documentclass[prb,twocolumn,balancelastpage,letterpaper,superscriptaddress]{revtex4-1}
\usepackage{times}

\usepackage{amsmath}
\usepackage{amssymb}
\usepackage{graphicx}
\usepackage{bm}
\usepackage{bbm}

\usepackage{color}
\usepackage{verbatim}

\usepackage{hyperref}

\newcommand{\eqn} {Eq.~}

\newcommand  {\ee}           {\mathrm{e}}
\newcommand  {\ii}           {\mathrm{i}}
\newcommand  {\ket}[1]       {\lvert#1\rangle}

\newcommand  {\spinup}       {\mathord{\uparrow}}
\newcommand  {\spindn}       {\mathord{\downarrow}}

\newcommand  {\nn}[1]        {\left\langle #1 \right\rangle}
\newcommand  {\nnn}[1]       {\left\langle\langle #1 \right\rangle\rangle}
\renewcommand{\vec}[1]       {\mathbf{#1}}                        
\renewcommand{\vr}           {\vec{r}}

\newcommand  {\ibar}         {{\bar{i}}}
\newcommand  {\sgn}          {\mathop{\mathrm{sgn}}}

\begin{document}

\title{Nontrivial topological states on a M\"obius band}

\author{W. Beugeling}
\affiliation{Max-Planck-Institut f\"ur Physik komplexer Systeme, N\"othnitzer Stra\ss e 38, 01187 Dresden, Germany}
\author{A. Quelle}
\affiliation{Institute for Theoretical Physics,  EMME$\Phi$, Utrecht University, Leuvenlaan 4, 3584 CE Utrecht, The Netherlands}
\author{C. \surname{Morais Smith}}
\affiliation{Institute for Theoretical Physics,  EMME$\Phi$, Utrecht University, Leuvenlaan 4, 3584 CE Utrecht, The Netherlands}

\date{\today}

\begin{abstract}
In the field of topological insulators, the topological properties of quantum states in samples with simple geometries, such as a cylinder or a ribbon, have been classified and understood during the last decade. Here, we extend these studies to a M\"obius band, and argue that its lack of orientability prevents a smooth global definition of parity-odd quantities such as pseudovectors. In particular, the Chern number, the topological invariant for the quantum Hall effect, lies in this class. The definition of spin on the M\"obius band translates into the idea of the orientable double cover, an analogy used to explain the possibility of having the quantum spin Hall effect on the M\"obius band. We also provide symmetry arguments to show the possible lattice structures and Hamiltonian terms for which topological states may exist in a M\"obius band, and we locate our systems in the classification of topological states. Then, we propose a method to calculate M\"obius dispersions from those of the cylinder, and we show the results for a honeycomb and a kagome M\"obius band with different types of edge termination. Although the quantum spin Hall effect may occur in these systems when intrinsic spin-orbit coupling is present, the quantum Hall effect is more intricate and requires the presence of a domain wall in the sample. We propose an experimental set-up which could allow for the realization of the elusive quantum Hall effect in a M\"obius band.
\end{abstract}

\maketitle


\section{Introduction}

Since topological insulators have been proposed theoretically \cite{KaneMele2005PRL95-14,KaneMele2005PRL95-22,BernevigEA2006} and observed experimentally \cite{KonigEA2007,HasanKane2010}, the notion of topology has received a growing interest from the condensed-matter physics community. The motivation for this interest resides on the great potential for technological use of systems exhibiting quantized currents that are robust against disorder, due to their topological protection. 

From a theoretical perspective, the topological protection is understood in terms of topological invariants. In two-dimensional electron gases in magnetic fields, the protected quantum Hall (QH) conductivity is governed by the TKNN integer,\cite{ThoulessEA1982} a specific case of a Chern number. For the quantum spin Hall (QSH) effect, the equivalent topological invariant is the spin Chern number.\cite{ShengEA2006} Classification of gapped free fermionic Hamiltonians in the presence of the fundamental discrete symmetries (time-reversal, particle-hole, and chiral) through the corresponding topological invariant is provided by the so-called ten-fold way.\cite{Kitaev2009,RyuEA2010}

The bulk-boundary correspondence\cite{Hatsugai1993PRB,*Hatsugai1993PRL} relates the Chern number of a bulk two-dimensional system and the number of edge modes in a finite system. Thus, the topological properties are conveniently determined by identifying the edge states in the dispersion of a quasi-one-dimensional system, i.e., a system which is infinite in one direction and finite in the other. This configuration could be thought of as a cylinder, where the translational symmetry in the infinite direction is encoded as periodic boundary conditions.\cite{HatsugaiEA2006,GoldmanEA2012,BeugelingEA2012PRB86-07}

Here, we raise the question as to whether the topological properties still exist when we consider a M\"obius band instead of a cylinder. The M\"obius band, which is an object with a topologically nontrivial geometry, is very appealing for physicists. Although the topology of the M\"obius band is fairly simple, the actual shape of a physical M\"obius band is far from trivial: a parametrization of this object that minimizes the bending and stretching of its constituent material can be found only through numerical calculation.\cite{StarostinVanDerHeijden2007} Topological effects of the ``twist'' on the quantum states in M\"obius ladders have been reported.\cite{ZhaoEA2009} Topological properties of the M\"obius geometry have recently been emulated electronically in capacitor-inductor networks.\cite{JiaEA2013preprint} The fascination is also fueled by the recent experimental progress in graphene nanoribbons. For graphene M\"obius ribbons, theoretical predictions of the electronic,\cite{YamashiroEA2004,JiangDai2008,CaetanoEA2009,GuoEA2009,WangEA2010,KorhonenKoskinen2014} magnetic,\cite{WakabayashiHarigaya2003,CaetanoEA2008} and thermal\cite{JiangEA2010} properties have been made, often in connection to structural and geometrical properties. Similar studies have been performed for boron nitride ribbons.\cite{AzevedoEA2012} The experimental realization of sufficiently wide M\"obius ribbons in these materials has not yet been reported, but should be considered possible, in the light of the successful realization of NbSe$_3$ M\"obius ribbons.\cite{TandaEA2002,*TandaEA2005} 

The most apparent difference between the cylinder and the M\"obius band is that the latter has only a single edge whereas the former has two. Differences in the \emph{electronic} topological properties are to be expected in view of the interpretation of the invariants in terms of edge currents. Furthermore, several ingredients in the topological analysis require the surface to be orientable.\cite{[{For an introduction to orientation and orientability, see, e.g.\ }][{}]Frankel2004book} For example, one cannot apply a uniform perpendicular magnetic field to a nonorientable surface as the M\"obius band, which inevitably leads to problems if we try to probe the QH invariant on this surface. On the other hand, the two spin degrees of freedom provide a possibility for the existence of the QSH effect on the M\"obius band.\cite{HuangLee2011}

In this work, we focus on the role of (non)orientability. Generally, it is impossible to define a pseudovectorial field smoothly on a nonorientable surface; the impossibility of applying a uniform perpendicular magnetic field on the M\"obius band is a specific example. A \emph{local} definition of such quantities is always possible, but the topology of the surface may prevent them to be defined \emph{globally} in a smooth way. In particular, the Chern number behaves in this way, and therefore requires the choice of an orientation, which cannot be done continuously on the M\"obius band (see Fig.~\ref{fig_qhe_magn}). The direct connection of the Chern numbers to Hall conductivity provides yet another argument against the existence of a QH effect on the M\"obius band.

The spin-$\tfrac{1}{2}$ degrees of freedom constitute two copies of each point of the base space, one for each spin component. In a cylindrical geometry, this construction yields two disconnected cylinders. For the M\"obius band, the spin space is its orientable double cover (ODC), the unique orientable manifold that has a two-to-one mapping to the base space. The ODC of the M\"obius band has a single connected component, unlike the cylindrical case.
The notion of the ODC is central to the existence of the QSH effect on the M\"obius band. Thus, we analyze the compatibility of the usual Hamiltonian terms, such as Zeeman effect, Rashba and intrinsic spin-orbit (SO) coupling, with the construction of the ODC.

The absence of the QH effect (and the usual Chern number) on the M\"obius band suggests that the usual classification of the topological invariants\cite{Kitaev2009,RyuEA2010} does not apply to the M\"obius band. The ``twist'' of the M\"obius band is interpreted in terms of a (glide) reflection symmetry, that alters the nature of the topological invariants. Recently, Chiu~\emph{et al.}\cite{ChiuEA2013} have extended the ten-fold way to spaces with reflection symmetries. The nature of the topological invariants depends on the presence of the reflection symmetry and whether it commutes or anticommutes with time-reversal and charge-conjugation symmetry (if present). Here, we analyze these commutation properties for the specific case of the M\"obius band, and we show that our findings are compatible with those of Ref.~\onlinecite{ChiuEA2013}.

The outline of the article is as follows. In Sec.~\ref{sect_topological_arguments}, we elaborate on the nonorientability of the M\"obius band together with the role of pseudovectorial quantities. We propose several nonuniform magnetic field configurations to generate local QH effects on the M\"obius band. The ODC construction is analyzed in connection to the QSH effect. In Sec.~\ref{sect_symmetries}, we discuss the symmetry properties of the lattice and of the terms in the Hamiltonian. We show that our results are compatible with the extended topological classification. We propose a way to compute band structures for the M\"obius band in Sec.~\ref{sect_band_structures}, and we show several examples of them. We conclude with a discussion, in Sec.~\ref{sect_discussion}, proposing a way to realize the magnetic field configurations discussed earlier.


\section{Topological arguments}%
\label{sect_topological_arguments}%

\subsection{Quantum Hall effect on the M\"obius band}
The QH effect arises in two-dimensional electron gases subjected to a perpendicular magnetic field. Its hallmark is that the Hall conductance $\sigma_{xy}$, defined by the in-plane current response to a  perpendicular (in-plane) voltage, is quantized in units of $e^2/h$. Both in experimental and theoretical analysis, the Hall measurement is generally performed on an orientable two-dimensional surface, such as a rectangle or a cylinder. These surfaces allow one to choose a single globally defined orientation. This choice fixes the sign of the Hall conductivity, which is therefore a well-defined quantity.

A nonorientable surface like the M\"obius band has the property that one cannot choose a global orientation. Orientations can be chosen locally, but it is not possible to connect them continuously on the whole surface. For any observable quantity to be well-defined, it must be independent of the choice of orientation, i.e.,  invariant under a change of orientation. We test this property by obtaining its behavior under the parity transformation $(x,y)\to(x,-y)$, since this mapping reverses orientation. Parity-odd quantities do not have an unambiguous definition: they change sign under a change of orientation, and thus they are orientation dependent.

The Hall conductance $\sigma_{xy}$, defined by $J_x = \sigma_{xy}E_y$ and $J_y = -\sigma_{xy}E_x$, where $J$ and $E$ are the current density and electric field, respectively, and the subscripts $x$ and $y$ label the components, is \emph{not} invariant under a parity transformation of space: this transformation inverts the sign of $\sigma_{xy}$. Thus, the Hall conductance is ill-defined on the M\"obius band. The Chern number, that is closely related to the Hall conductivity, shares this property. It is defined as the integral of the Berry curvature, which in turn is the curl of the Berry connection. The definition of the curl requires a choice of orientation, so that the definition is ambiguous on a nonorientable surface.\footnote{The Brillouin zone is nonorientable, because it inherits its symmetry properties under parity transformation of the momentum space from those of the original space.}

A similar reasoning may be used for the perpendicular magnetic field that conventionally generates the QH effect. The magnetic field is a \emph{pseudovector}, meaning that it changes sign under a parity transformation. As a consequence, one cannot apply a uniform magnetic field to a M\"obius band. It remains possible to apply \emph{local} magnetic fields, i.e., fields that depend on the spatial coordinates.


\begin{figure}
  \includegraphics[width=84mm]{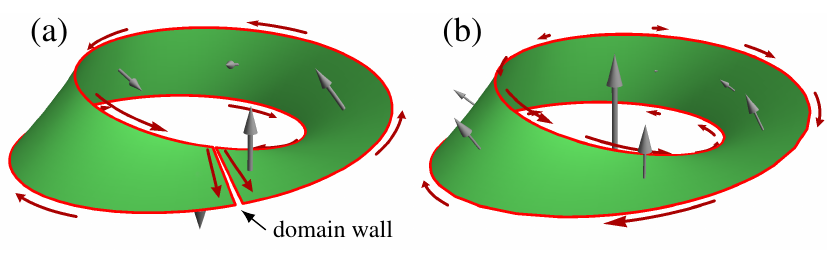}
  \caption{(Color online) Two configurations of the M\"obius band with magnetic fields (gray arrows) and edge currents (red arrows). (a) M\"obius band with a magnetic field perpendicular to the surface and of constant magnitude. At one transversal domain wall it changes sign. There are currents through the bulk at the domain wall. (b) M\"obius band subjected to a uniform magnetic field in the embedding space, indicated by the large arrow. The component perpendicular to the surface, indicated by the arrows on the surface, varies smoothly (cosine-like). The bulk currents are not indicated.}
  \label{fig_qhe_magn}
\end{figure}


The piecewise definition of the Hall conductivity provides the possibility for currents along the edge even on a nonorientable surface. Let us consider two classes of configurations on the M\"obius band. First, assume a perpendicular magnetic field depending on the $x$ (longitudinal) coordinate. In this case, we locally have two counterpropagating edge modes. The currents propagate clockwise on half of the edge and counterclockwise on the other half. In Fig.~\ref{fig_qhe_magn}, we show two examples of such configurations. In panel (a), the magnetic field is everywhere perpendicular to the surface and constant, except at one line spanning the band from one edge to the other. At this line, the direction of the magnetic field inverts, and for this reason we name it a domain wall. At the domain wall, currents flow through the bulk.\cite{HuangLee2011} In Fig.~\ref{fig_qhe_magn}(b), the magnitude of the perpendicular magnetic field is equal to the perpendicular component of a ``background'' magnetic field, i.e., a uniform magnetic field in the embedding space. In the coordinate system of the surface, the magnetic field dependence is cosine-like; thus, it is smooth except where the orientation is discontinuous. Transverse currents that exist everywhere on the surface destroy the quantization of the edge currents. In both examples, the average edge current, defined as the integral of the current over the full edge divided by the length, is zero.

We may also probe the Hall conductivity in a different manner. As explained in Ref.~\onlinecite{Girvin1999lecturenotes}, a quantum Hall fluid in the Corbino geometry (an annulus) may be pierced through the central hole by an infinitely thin solenoid. As the flux inside the solenoid is increased adiabatically by one flux quantum $h/e$, a circular current is generated inside the annulus, which in turn induces a Hall voltage. In this situation, the electric charge on the inner and outer boundary is $\pm\sigma_{xy}h/e$. Thus, a charge of $ne$ on the boundary indicates that the Hall conductivity is equal to (the integer) $n$ times the conductivity quantum $e^2/h$. When we repeat this thought experiment for the M\"obius band, with a solenoid flux tube through the central hole, the adiabatic flux change induces a circular current. However, as opposed to a separate inner and outer boundary in the Corbino geometry, here there is only one boundary. Conservation of the boundary charge requires that it vanishes, and thus we infer that the Hall conductivity does so as well.


\begin{figure}
  \includegraphics[width=84mm]{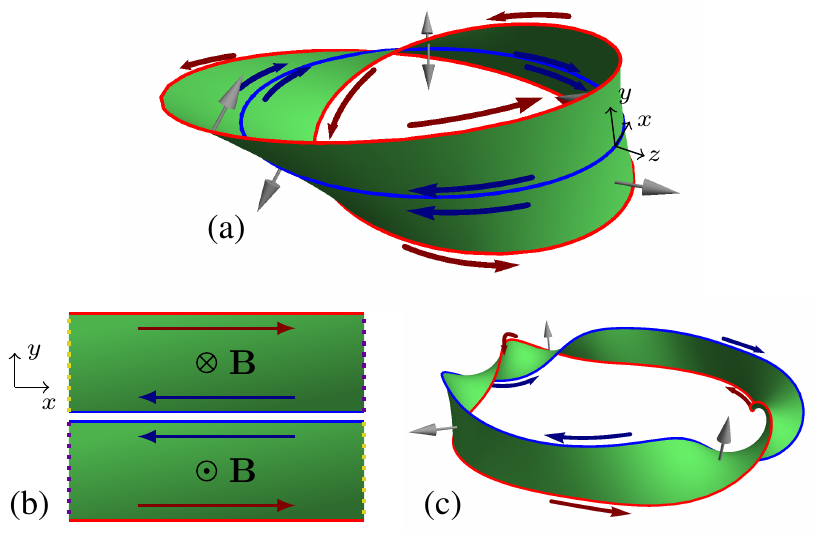}
  \caption{(Color online) QH state on the M\"obius band with a longitudinal cut. (a) The M\"obius band is cut in two by the center line (blue). Magnetic fields (gray arrows) have opposite directions on both sides of the center line. (b) Cut-open model of (a). The pairs of yellow and purple edges should be identified in order to obtain the M\"obius band shown in (a).  The red and blue arrows along the edges indicate the propagation direction of the edge currents in the presence of the magnetic field ${\bf B}$. (c) The M\"obius band cut at the center line yields a $4\pi$ twisted ribbon. The edge currents and magnetic fields are indicated as in (a).}
  \label{fig_qhe_cut}
\end{figure}


Secondly, we consider a magnetic field that is uniform in the longitudinal direction, but dependent on the transversal coordinate $y$. The parity transformation property requires that the perpendicular component $B_z$ of the magnetic field satisfies $B_z(-y)=-B_z(y)$. In particular, on the center line $y=0$ (i.e., halfway between two edges, seen locally), the field must vanish. One particularly interesting example is $B_z(y)=B_z\sgn(y)$, see Figs.~\ref{fig_qhe_cut}(a) and (b). In this configuration, Hall currents propagate on the edges in one direction, and on the center line in the opposite direction. Without affecting the topological transport properties, we can cut the M\"obius band at the discontinuity on the center line. The resulting surface is homeomorphic (topologically equivalent) to a ribbon with a $4\pi$ twist, see Fig.~\ref{fig_qhe_cut}(c). This manifold is orientable and has two separate (disconnected) edges. Thus, this configuration is equivalent to a cylinder with a uniform perpendicular magnetic field, with oppositely propagating edge currents on both edges. We note that one of the edges is the original edge of the M\"obius band, the other is the center-line cut.

If we would perform a Hall measurement on half of the ribbon, by probing the area between the edge and the center line, we would find a nonzero Hall conductivity. If we probe the M\"obius band as a whole, we find a zero \emph{total} Hall conductivity, because the contributions from  either side of the center line cancel. The zero total Hall current is intuitive from the fact that the two edge currents locally copropagate [see the red arrows in Fig.~\ref{fig_qhe_cut}(a)]. The solenoid thought experiment described above has an interesting outcome in this configuration: By adiabatically increasing the flux in the solenoid through the central hole, one generates a circular current through the whole band. The current induces the charges $2ne$ on the center line and $-2ne$ on the edge, associated to the Hall conductivity of magnitude $ne^2/h$ and opposite signs on opposite sides. This example shows that one should make a clear distinction between local (piecewise) Hall currents and the total Hall current. Thus, the statement that the M\"obius band does not admit a total nonzero Hall current remains valid, even with a local field configuration.

\subsection{Quantum spin Hall effect on the M\"obius band}
In conventional systems, the QSH effect can be understood from treating spin up and spin down particles being subjected to equally large but opposite magnetic fields. Although uniform magnetic fields and nonzero total Hall currents are forbidden on the M\"obius band, QSH-like states do exist there.\cite{HuangLee2011}

Spin refers to two internal degrees of freedom at each point in space. Locally, there are two copies of the space, usually labeled with spin up and spin down, but we may consider more generally any two spins connected by time-reversal symmetry. In the description of the QSH effect on a cylinder, the two copies are defined globally. There is a spin up and a spin down cylinder, and there is no continuous spatial transformation that can turn a spin up into a spin down and vice versa. In theory, the same construction could be used to make two disconnected copies of the M\"obius band, but, because neither of the two copies admits a nonzero Hall current, this construction cannot explain the existence of the QSH-like state.


\begin{figure}
  \includegraphics[width=84mm]{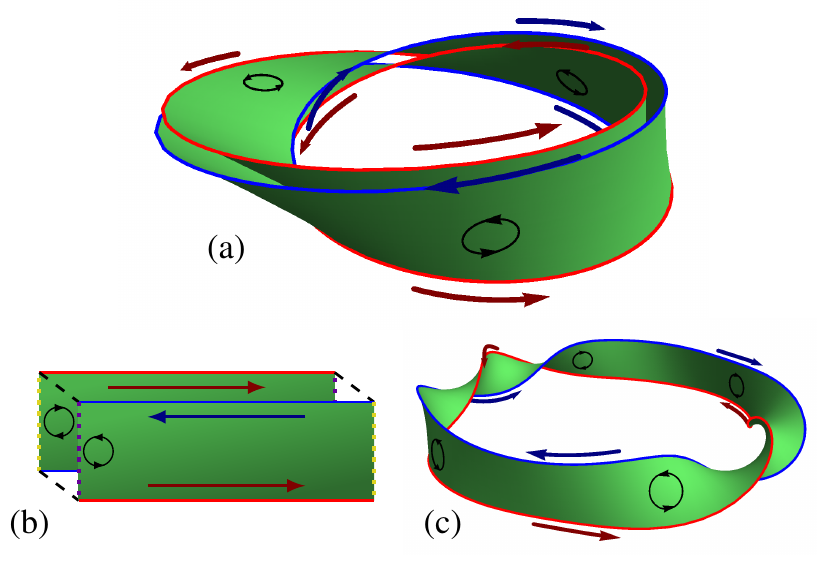}
  \caption{(Color online) QSH-like state on the M\"obius band with the double-covering description. (a) Double covering of the M\"obius band where the two copies are separated by a small distance. We indicate the edge currents as in Fig.~\ref{fig_qhe_cut}. The circular arrows on the surface determine the ``chirality'' of the Haldane flux. (b) Cut open version of the double cover. The chirality of the Haldane fluxes and the edge currents are indicated. We recall that the two copies spatially coincide in reality and that the offset is for mere illustrational purposes. (c) The double cover is homeomorphic to the $4\pi$ twisted band. Again, we indicate the chirality of the Haldane flux and the edge currents. The edge-state configuration is identical to the one shown in Fig.~\ref{fig_qhe_cut}, thus proving their topological equivalence.}  
  \label{fig_qshe_double_cover}
\end{figure}


An alternative way to assign locally two copies of the M\"obius band is known as the \emph{orientable double cover} (ODC). The ODC of a manifold $M$ constitutes the manifold $D$, consisting of points $(\vr,o)$ for each $\vr\in M$ and $o$ one of the two orientations, together with the two-to-one covering map from $D$ to $M$ defined by $(\vr,o)\mapsto \vr$. This cover exists and is unique for any nonorientable surface. (For orientable manifolds, the double covering map also exists and is trivial.) The two local copies at each point $\vr$ are interpreted as the two spin components. As shown in Fig.~\ref{fig_qshe_double_cover}, the double cover of the M\"obius band is homeomorphic to the $4\pi$ twisted ribbon, an orientable surface.

Suppose that we subject the double cover (or equivalently, the $4\pi$ twisted ribbon) to a Haldane-like flux configuration on the honeycomb lattice, which is characterized by a vanishing total flux, but which still generates a QH effect.\cite{Haldane1988} The Haldane flux has a well-defined chirality, defined by the direction of the edge currents it induces. We stress that these notions are unambiguous on the ODC because it is orientable. Then, by virtue of the covering map, the two components at each point on the M\"obius band are subjected to two Haldane fluxes of opposite chirality. This configuration essentially defines the intrinsic SO coupling term of Kane and Mele,\cite{KaneMele2005PRL95-14,KaneMele2005PRL95-22} that induces the QSH effect: The two edge modes at each edge counterpropagate, and the edge modes of matching components at both sides of the ribbon counterpropagate as well, see Fig.~\ref{fig_qshe_double_cover}. An important difference with the QSH state in the cylindrical case is that the spin labels cannot be assigned globally, because the double cover has only one connected component. Indeed, a translation once around the central hole in the M\"obius band, i.e., such that the longitudinal coordinate is the same as before, maps spin up to spin down and vice versa. The spin flip connected to this single rotation has been observed in measurements of the time-resolved dynamics in RF circuits emulating the M\"obius band.\cite{JiaEA2013preprint}

By comparing Figs.~\ref{fig_qhe_cut}(c) and \ref{fig_qshe_double_cover}(c), we find that the QSH effect on the M\"obius band is topologically equivalent to the configuration with a longitudinal cut. In both cases, the total Hall conductivity vanishes. However, there is an important difference in the interpretation. In the spinful case, there are two edge states, whereas in the spinless case there is only one; the other one is ``hidden'' in the bulk along the center line. Only in the spin model, one can speak of a spin Chern number and a nonzero QSH conductivity (the sign of which may be fixed from the mapping to the $4\pi$ ribbon).

\section{Hamiltonian and lattice symmetries}%
\label{sect_symmetries}%

\subsection{Lattice symmetries}%
\label{sect_symmetries_lattice}%
Not all lattices can be placed on the M\"obius band in a regular manner: When ``glueing'' a ribbon into a M\"obius band, the result should not show any signs of a ``cut'' where the glueing has taken place. In other words, the lattice on the M\"obius band shows nontrivial (discrete) translational symmetries. For this reason, the original lattice must have at least one axis of reflection symmetry. Fortunately, the lattice structures most commonly studied in the context of the quantum (spin) Hall effect have this property. Even more interestingly, these lattices generally have two inequivalent axes of symmetry, which means that a ribbon can be glued into a M\"obius band in at least two different ways.

Typically, when one studies the edge states on a cylinder, one considers a ribbon in the $x$ direction and one identifies all sites related by the discrete translation $(x,y)\mapsto(x+a,y)$, where $a$ is such that $(x+a,y)$ is a point on the lattice if and only if $(x,y)$ is. For a M\"obius band, we identify sites related by the glide reflection $G:(x,y)\mapsto(x+a,-y)$. The reflection allows only two possible edge directions, zigzag and armchair, because the edge must be parallel to one of the symmetry axes. This should be contrasted to the cylindrical case where the edge can be in any direction.


\begin{figure}[t]
  \includegraphics{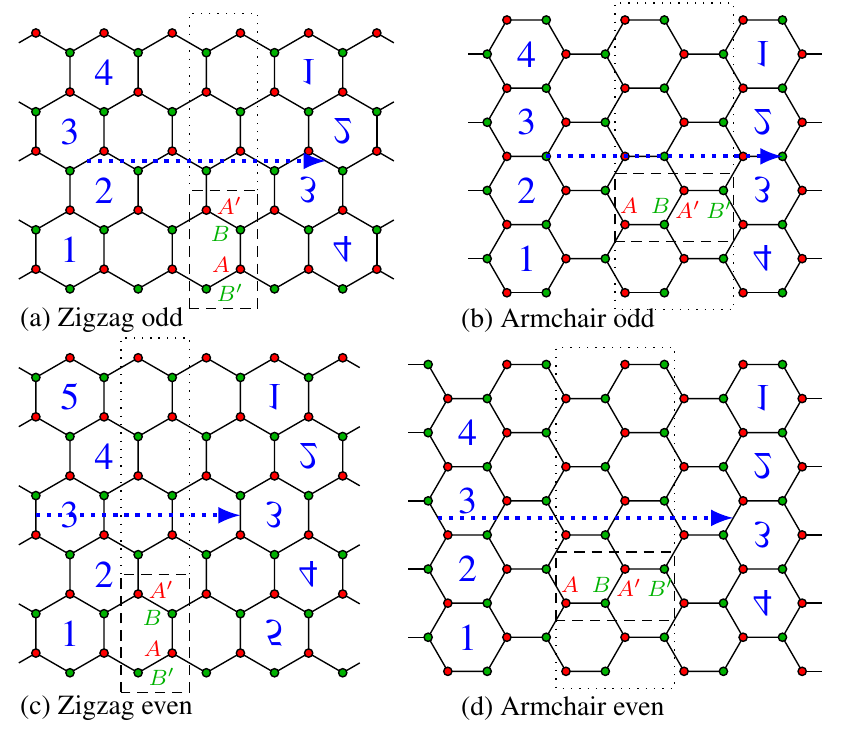}
  \caption{(Color online) Construction of honeycomb M\"obius bands from a ribbon. The left and right edge are ``glued together'' such that the blue numbers on the left- and right-hand sides coincide. This construction is possible due to the invariance under the glide reflection, with a translation indicated by the dotted arrow and the reflection perpendicular to this arrow. The dashed rectangle indicates a bulk  unit cell with the four sublattice labels. The dotted rectangle is a ribbon unit cell. The four panels show zigzag and armchair, with ``even'' and ``odd'' widths.}
  \label{fig_honeycomb_lattice}
\end{figure}


Let us discuss the honeycomb lattice as a first example. In Fig.~\ref{fig_honeycomb_lattice}, we show zigzag-edge and armchair-edge ribbons with different widths, together with a glide reflection that defines the identification (``glueing'') that leads to a M\"obius band. The reflection part of the glide reflection dictates that the elementary (bulk) unit cell is rectangular. The lattice vectors are parallel to the coordinate axes and hence perpendicular to each other. For both the zigzag and the armchair bands, we therefore need to choose a unit cell with four sites. In comparison, for the cylindrical case, the lattice vectors need not be perpendicular, and can therefore be described with a two-site unit cell in the zigzag-edged case.\cite{HatsugaiEA2006,BeugelingEA2012PRB86-07}

For computing a ribbon dispersion, we treat all sites in a rectangular region spanning from bottom to top edge, that we call the ribbon unit cell. For the glueing, we distinguish two cases where the transverse direction can be described by an integer number of bulk unit cells or not, which we call ``even'' and ``odd'', respectively. These labels refer to the number of $A$ and $B$ sites in the transverse direction, being $2w$ and $2w-1$ in the even and odd case, respectively, where $w$ is an integer that denotes the width of the ribbon. (The total number of sites in the transverse direction is $4w$ and $4w-2$ respectively.) Numbering the bulk unit cells $i=1,\ldots,w$ from top to bottom (where there is only ``half'' a unit cell for $i=w$ in the odd case), the identification induced by the glide reflection $G$ can be characterized as
\begin{equation}\label{eqn_sublattice_mapping}
  \begin{aligned}
    (A_i,B_i,A'_i,B'_i) &\leftrightarrow(B_\ibar,A_\ibar,B'_\ibar,A'_\ibar)&&\text{(zigzag, even)}\\
    (A_i,B_i,A'_i,B'_i) &\leftrightarrow(B'_\ibar,A'_{\ibar-1},B_{\ibar-1},A_\ibar)&&\text{(zigzag, odd)}\\
    (A_i,B_i,A'_i,B'_i) &\leftrightarrow(A'_\ibar,B'_\ibar,A_\ibar,B_\ibar)&&\text{(armchair, even)}\\
    (A_i,B_i,A'_i,B'_i) &\leftrightarrow(A_\ibar,B_\ibar,A'_{\ibar-1},B'_{\ibar-1})&&\text{(armchair, odd)}
  \end{aligned}
\end{equation}
where $(A_i,B_i,A'_i,B'_i)$ denote the four sites of unit cell $i$, and $\ibar$ is shorthand for $w+1-i$. We note the differences in the invariance of the sublattice labeling. The four possible actions of the glide reflection are generated by the exchanges $(A,A')\leftrightarrow(B,B')$ and $(A,B)\leftrightarrow(A',B')$, forming a $\mathbb{Z}_2\times\mathbb{Z}_2$ group structure. The transformation $(A,B)\leftrightarrow(A',B')$ preserves the ``color'' of the sites, where the pairs $(A,A')$ and $(B,B')$ each have a single color (see Fig.~\ref{fig_honeycomb_lattice}). The reflection is color preserving for the armchair and color inverting for the zigzag case.


\begin{figure}[t]
  \includegraphics{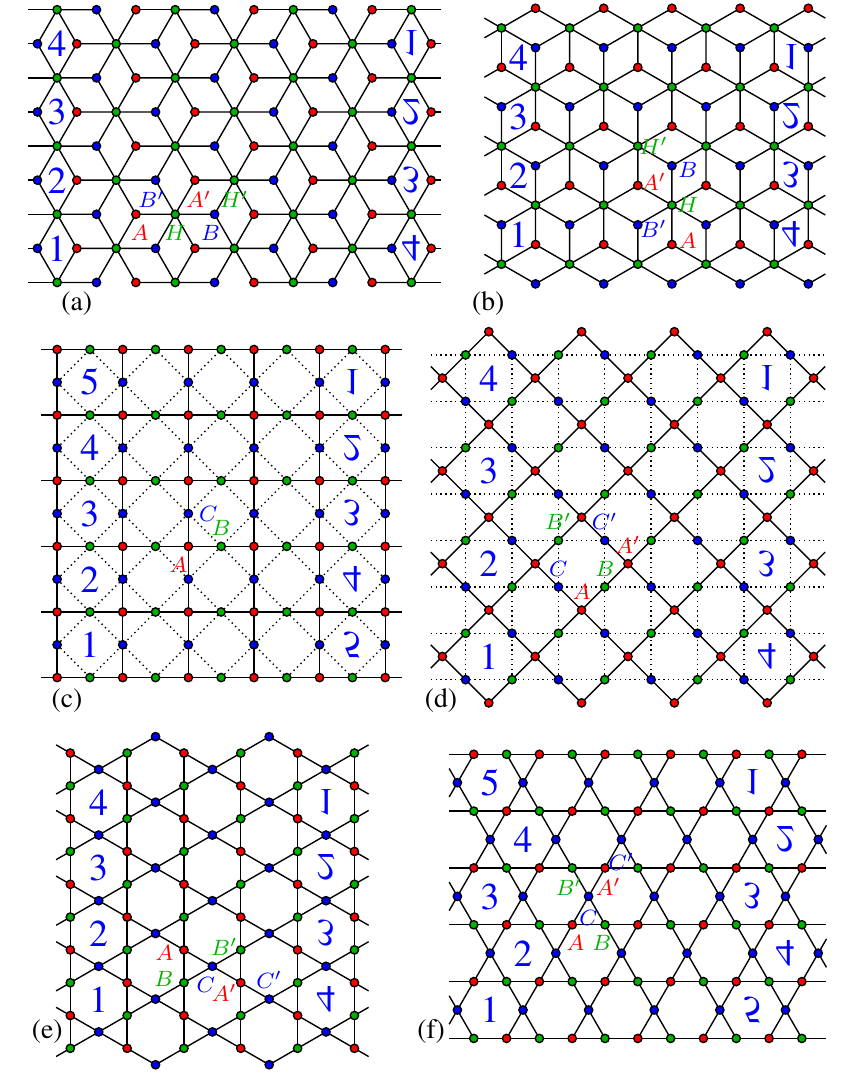}
  \caption{(a,b) The dice lattice. (c,d) The Lieb lattice. (e,f) The kagome lattice. The letter labels indicate a possible choice of the unit cell with sublattice labeling. The dotted lines indicates the next-nearest neighbor (intrinsic SO) coupling. The numbers on the left and right of each lattice indicate how a strip with this lattice structure should be ``folded'' into a M\"obius band: The left and right edge should be glued together such that the corresponding number symbols coincide.}
  \label{fig_lattices}
\end{figure}


This discussion also applies to other lattice structures known for showing topological effects, e.g., the dice,\cite{BerciouxEA2011} Lieb,\cite{WeeksFranz2010,GoldmanEA2011} and kagome\cite{GuoFranz2009} lattices, see Fig.~\ref{fig_lattices}. These examples all have three atoms per unit cell in the bulk lattice. The dice lattice could be considered as a honeycomb lattice with one extra site in each unit cell. For the M\"obius band, a six-site unit cell must be taken, but the action of the glide reflection on the sublattice structure is analogous to that of the honeycomb lattice. The Lieb lattice has two inequivalent symmetry axes under an angle of $45^\circ$. The M\"obius band with a straight edge requires only three sites per unit cell, whereas the one with a zigzag edge requires six, see Figs.~\ref{fig_lattices}(c) and (d), respectively. The kagome lattice has the peculiarity that there is no ``color'' preserving glide reflection. In other words, the kagome lattice is chiral: The left-handed and right-handed version cannot be mapped onto each other by translation and rotation.

\subsection{Symmetries of the Hamiltonian}

The construction of the M\"obius band requires not only the lattice to be invariant under the glide reflection, but also the Hamiltonian of the system must satisfy this property. For the Hamiltonian terms involving spin, such as the Zeeman term, as well as the intrinsic and the Rashba SO coupling, we need to map the two spin components into the two components of the double cover of the M\"obius band. Recall that a local mapping is always possible, but the possibility of performing this mapping for the Hamiltonian term depends on whether it can be defined globally in a continuous way. Thus, the Hamiltonian must be invariant under the transformation $T=G\sigma_x$ defined by the simultaneous action of the aforementioned glide reflection $G$ and a ``spin flip''; the latter being equivalent to an exchange of the two components of the double cover. In other words, the Hamiltonian must commute with the action of $T$. This property may be seen as analogous to the cylindrical case, where the Hamiltonian is invariant under translation by the periodicity. Because applying $T$ twice defines a pure translation, the allowed Hamiltonian terms for the M\"obius band are a subset of those of the cylinder. This observation motivates why we discuss several terms that are commonly studied in the context of the quantum (spin) Hall effect on the cylinder,\cite{GoldmanEA2012, BeugelingEA2012PRB86-07} and derive whether they are symmetric under $T$ as well.

First, ordinary nearest-neighbor hopping in absence of a magnetic field is invariant under a glide reflection and can therefore appear in the tight-binding Hamiltonian for the M\"obius band. If the hopping would be subject to a magnetic flux through the lattice, it would pick up complex phase factors $\ee^{\ii\theta_{jk}}$, as
\begin{equation}\label{eqn_hnn}
H_\mathrm{NN} = -t \sum_{\nn{j,k}} \ee^{\ii\theta_{jk}} c^\dagger_j c_k,
\end{equation}
where the sum is over nearest-neighbor sites $j$ and $k$, $t$ is the hopping amplitude, and $c_i$ ($c^\dagger_i$) denotes the annihilation (creation) operator on site $i$. The glide reflection leaves $\theta_{jk}$ invariant.  The reflection does invert rotational sense, i.e., it transforms clockwise to counterclockwise rotation and vice versa, so that the flux through a loop, given as the sum of the hopping phases in counterclockwise direction, flips its sign. Configurations where $B_z(-y)=-B_z(y)$ are possible if we choose $\theta_{j'k'}=\theta_{jk}$ where the reflection maps sites $j$ and $k$ onto $j'$ and $k'$, respectively. 

The Zeeman term $H_\mathrm{Z} = -t_\mathrm{Z} \sum c_j^\dagger \sigma_z c_k$ anticommutes with transformation $T$, and is therefore not allowed on the M\"obius band. Alternatively, one could argue that the Zeeman term requires a global ``labeling'' of the spin up and spin down components, which we have proven not to exist because the double cover consists of one connected component.

A staggered sublattice potential, as used by, e.g., Kane and Mele,\cite{KaneMele2005PRL95-14} can be applied only if it is compatible with the reflection properties. In the terminology of Sec.~\ref{sect_symmetries_lattice}, the reflection must be ``color preserving''. For honeycomb ribbons, this property is satisfied only for the armchair configuration, as shown by \eqn\eqref{eqn_sublattice_mapping}.

The SO terms are characterized by a coupling between the spin degrees of freedom and the momentum of the charge carriers. The Rashba SO term on a lattice can be written as
\begin{align}\label{eqn_hrso}
  H_\mathrm{R} 
  &= -\ii t_\mathrm{R}\sum_{\nn{j,k}}c^\dagger_j(\sigma_x d^y_{jk}-\sigma_y d^x_{jk})c_k\\
  &= t_\mathrm{R}\sum_{\nn{j,k}}\left[ c^\dagger_{j,\spinup}(d^x_{jk}-\ii d^y_{jk})c_{k,\spindn}+c^\dagger_{j,\spindn}(d^x_{jk}+\ii d^y_{jk})c_{k,\spinup}\right],\nonumber
\end{align}
where $(d^x_{jk},d^y_{jk})$ denotes the vector from site $j$ to $k$, and $c_k=(c_{k,\spinup},c_{k,\spindn})$ is a spinor. The latter form of $H_\mathrm{R}$ in \eqn\eqref{eqn_hrso} shows the invariance under the combined glide reflection [which acts as $(d^x,d^y)\mapsto(d^x,-d^y)$] and spin flip. The intrinsic SO coupling acts as a next-nearest neighbor term,
\begin{align}\label{eqn_hiso}
  H_\mathrm{I}
  &= -\ii t_\mathrm{I} \sum_{\nnn{j,k}}\nu_{jk}c^\dagger_j \sigma_z c_k\\
  &=  -\ii t_\mathrm{I} \sum_{\nnn{j,k}}\nu_{jk}(c^\dagger_{j,\spinup}c_{k,\spinup} - c^\dagger_{j,\spindn}c_{k,\spindn}),\nonumber
\end{align}
where $\nu_{jk}$ is the sign of $d^x_{jl}d^y_{lk}-d^y_{jl}d^x_{lk}$ with $(d^x_{jl},d^y_{jl})$ and $(d^x_{lk},d^y_{lk})$ the nearest-neighbor vectors that connect sites $j$ and $k$ via an intermediate site $l$. 
The invariance of this term follows from the fact that $\nu_{jk}$ changes sign under reflection, while spin flip also acts as a sign change. We thus find that both the Rashba and intrinsic SO coupling terms are allowed to appear in the tight-binding Hamiltonian on the M\"obius band. Real NNN hopping \cite{[{See, e.g., }][{}]BeugelingEA2012PRB86-19} is closely related to intrinsic SO, but does not contain the spin flip that is essential for the invariance under the reflection. Hence, the real NNN hopping term is forbidden on the M\"obius band.

On an armchair-edged honeycomb ribbon, the Hamiltonian proposed in Ref.~\onlinecite{HatsugaiEA2006} with nearest-neighbour hopping \eqn\eqref{eqn_hnn} together with a third-neighbour hopping $H'=-t'\sum_{j,k}c^\dagger_jc_{k}$ for site pairs $(j,k)$ along the diagonals of the hexagons parallel to the $x$ axis, is glide-reflection symmetric (assuming no magnetic flux) and can therefore be defined on the M\"obius band. Interestingly, the parameter $t'$ can be varied continuously, which effectively changes the lattice geometry between the square lattice ($t'/t=1$), the honeycomb lattice ($t'/t=0$), and the $\pi$-flux lattice ($t'/t=-1$).\cite{HatsugaiEA2006} Similarly, a zigzag-edge ribbon with third-neighbour hopping parallel to the $y$ axis can be tuned to the same square and $\pi$-flux lattices.

\subsection{Topological classification}
Having argued based on orientability that the M\"obius band does not admit QH, but does admit QSH-like states, one could ask the question where the system would fit in the classification of topological insulators.\cite{Kitaev2009,RyuEA2010} In this classification, the nature of the topological invariant is determined by the symmetry properties under the time-reversal, particle-hole, and chiral symmetry operations. Recently, it has been shown that if in addition the system is reflection symmetric, the topological invariant can be different.\cite{ChiuEA2013} The topological invariant depends on whether the reflection operator commutes or anticommutes with the time-reversal, particle-hole, and/or chiral symmetry transformations, whichever is present.

In the classification table of Ref.~\onlinecite{ChiuEA2013}, the relevant dimensionality is $2$, and the symmetry class is either A (no time-reversal, no particle-hole, and no chiral symmetry) or AII (time-reversal symmetry only, and the time-reversal operator $\Theta$ squares to $-1$). For the A class in two dimensions the $\mathbb{Z}$ topological invariant that classifies the QH effect in absence of the reflection symmetry is turned into a trivial invariant for a reflection invariant system. This result is consistent with the idea that a uniform QH effect cannot exist in a reflection symmetric system like the M\"obius band. We note that the classification does not encompass local field configurations, and as such does not contradict the existence of a local QH effect in the example with the longitudinal cut of Sec.~\ref{sect_topological_arguments}.

For reflection symmetric systems in the AII class, the topological invariant can be either trivial or $\mathbb{Z}_2$ depending on whether the transformation $T=G\sigma_x$ commutes or anticommutes with time-reversal $\Theta$.\cite{ChiuEA2013} In our case, the time-reversal symmetry operator acts as
\begin{equation}\label{eqn_time_reversal}
  \Theta = \ee^{-\ii\pi\sigma_y/2}K = -\ii\sigma_y K,
\end{equation}
where $K$ denotes complex conjugation. Writing $T=XP\sigma_x$, where $X$ is the translation part of the glide reflection, $P$ is the parity transformation mapping $(x,y)$ into $(x,-y)$, and $\sigma_x$ encodes the spin flip. Thus, anticommutation follows by virtue of
\begin{equation}
  T^{-1}\Theta T=-\ii PX^{-1}\sigma_x \sigma_y K \sigma_x XP=\ii\sigma_y K = -\Theta.
\end{equation}
As a result, we find that the topological invariant is of $\mathbb{Z}_2$ type, identical to the one that characterizes the QSH effect in non-reflection-symmetric systems. This result does not mean that the topological state in the M\"obius band necessarily is the QSH state, but only that it is similar in nature. In this case, the interpretation of the spin Chern number in terms of spin up and down is not possible, contrarily to the usual QSH state.

The topological classification of Ref.~\onlinecite{ChiuEA2013} allows for a generalization of the results presented here, such as topological superconductors in two dimensions and topological insulators in higher dimensions. We refrain from further discussion in this direction as we consider it to be outside the scope of this article.


\section{Band structures}%
\label{sect_band_structures}%

\subsection{Method}%
\label{subsec_band_structure_method}%

In this section, we show the dispersions for a M\"obius band subjected to a magnetic flux with a domain wall along the longitudinal center line, as well as for the case of intrinsic SO coupling, cf.\ Figs.~\ref{fig_qhe_cut}(a) and \ref{fig_qshe_double_cover}(a), respectively. 
The computation of the band structures and edge-state dispersion on the M\"obius band is done by diagonalizing the corresponding Hamiltonian on a cylinder and halving the number of degrees of freedom, as described below. This procedure is possible because the Hamiltonian commutes with the transformation $T$ defined by composition of glide reflection and spin flip. Each eigenstate is doubly degenerate: each eigenstate $\ket{\psi}$ of the Hamiltonian has a partner eigenstate $T\ket{\psi}$ with the same eigenvalue. Naturally, the eigenstates could be labeled by their eigenvalue of $T$, which takes the values $+1$ and $-1$ for $T$-even and $T$-odd states, with eigenspaces spanned by $(1/\sqrt{2})(\ket{\psi}+T\ket{\psi})$ and $(1/\sqrt{2})(\ket{\psi}-T\ket{\psi})$, respectively. Whereas the Hilbert space for the cylinder contains $T$-even as well as $T$-odd states, the latter are unphysical in the M\"obius band  because all states here must be invariant under $T$. Thus, the spectrum of the M\"obius band is found by discarding all $T$-odd states.

The computation of the dispersions proceeds as follows. As a starting point, we take the Hamiltonian for the cylindrical case, including the spin degrees of freedom. For the spin down components, we substitute $y \to -y$ while keeping $k_x$. (Here, it is required that the spin components are uncoupled.) Instead of diagonalizing the resulting M\"obius band Hamiltonian $H_\mathrm{MB}$, we diagonalize $H_\mathrm{MB}(\alpha) = H_\mathrm{MB}+\alpha T$, where $\alpha$ is a parameter larger than the difference between the maximum and minimum energy of the dispersion of $H_\mathrm{MB}$. This method automatically resolves the degeneracy between any pair of degenerate eigenstates, without having to explicitly diagonalize $T$ for each two-fold degenerate eigenspace of $H_\mathrm{MB}$. The (uniquely defined) $T$-even and $T$-odd states of the pair at energy $E_n$  have $(H_\mathrm{MB}+\alpha T)$-eigenvalues $E_n+\alpha$ and $E_n-\alpha$, respectively. By choosing $\alpha$ to be large, the $T$-even and $T$-odd states become completely separated, and the $T$-odd states can then be discarded straightforwardly. The eigenstates found with this procedure do not depend on the value of $\alpha$, by virtue of the fact that  $H_\mathrm{MB}$ and $T$ share a common basis of eigenstates.

\subsection{Quantum Hall effect on a M\"obius band with a longitudinal domain wall or cut}


\begin{figure}[t]
  \includegraphics[width=84mm]{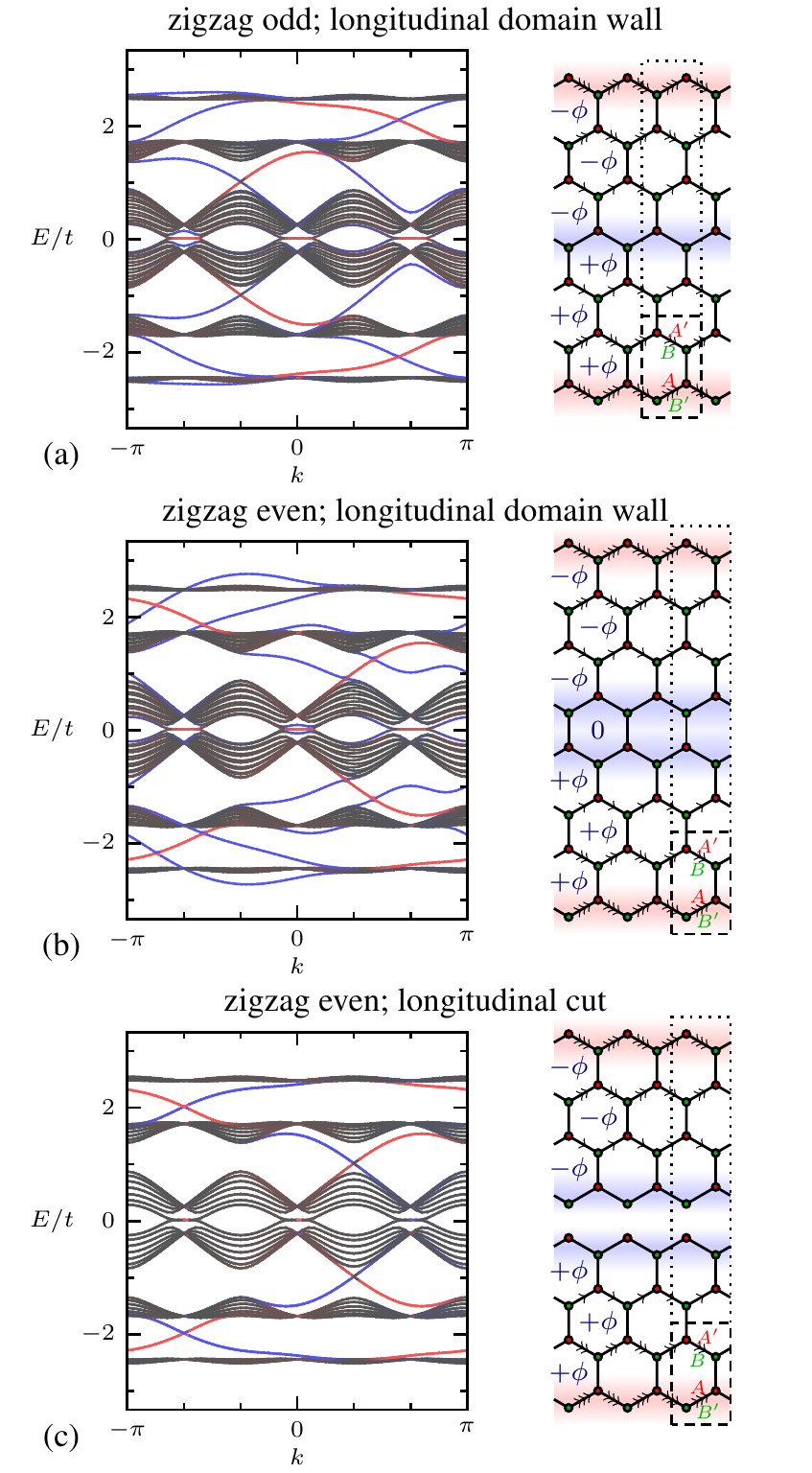}
  \caption{(Color online) Dispersions of the M\"obius band of the honeycomb lattice with zigzag edges, subjected to a magnetic flux of $1/3$ flux quantum per hexagon, with opposite signs on both sides of the center line. (We set the bond length to unity.) The edge states localized on the center line are shown in blue, and those on the M\"obius edge in red. On the right-hand side, we sketch the lattices with the flux phases assigned on the bonds: $n$ arrows indicate a phase factor $\ee^{\ii\theta_{jk}}=\ee^{\ii\pi n \phi}$ for the hopping in the direction of the arrows, where $\phi$ is the flux in units of the flux quantum $h/e$. The dispersions and flux configurations are shown for (a) a longitudinal \emph{domain wall} in an odd-sized zigzag ribbon, (b) a longitudinal \emph{domain wall} in an even-sized ribbon, and (c) a longitudinal \emph{cut} in an even-sized ribbon.}
  \label{fig_disp_qhe}
\end{figure}


In Fig.~\ref{fig_disp_qhe}(a) and (b), we show the dispersions for the M\"obius band subjected to a magnetic flux with a longitudinal domain wall. For illustrational purposes, we choose $1/3$ of a flux quantum per hexagon. In order to compute this dispersion, we have taken the ordinary Hamiltonian for a cylindrical ribbon, and assigned hopping phases $\theta_{jk}$ (see Fig.~\ref{fig_disp_qhe}) such that the flux through the bottom half and the top half have opposite signs. The hexagons on the center line are not subjected to flux. The hopping strengths for the bonds crossing the center line are equal to $t$, i.e., equal to the magnitude of all other hopping amplitudes. This configuration of a longitudinal \emph{domain wall} should be contrasted to that of a longitudinal \emph{cut}, where the bonds across the central line are cut, i.e., their hopping amplitudes are null. The resulting dispersion for the latter case is shown in Fig.~\ref{fig_disp_qhe}(c).

We interpret the resulting dispersions by studying the edge states in the bulk gap at $E/t=1$. (The other bulk gaps show qualitatively similar results.) In the domain-wall configuration [Figs.~\ref{fig_disp_qhe}(a) and (b)], we find edge states on the edge of the M\"obius band, shown in red. They are two-fold degenerate, because of the extra degrees of freedom included in order to describe the M\"obius geometry. Alternatively, one could explain the number of two by recalling that at each $x$ (longitudinal) coordinate of the M\"obius band, there are two edges. The edge states at the center line (colored blue) propagate in the opposite direction. This pair of edge states is not degenerate, because they overlap and hybridize, which causes an energy splitting lifting the degeneracy. As a consequence, the spectrum has no vertical axis of reflection symmetry ($k\to-k$). The combination of the magnetic flux (being chiral) and the different natures of the edge and the longitudinal domain wall is the cause of this symmetry breaking. 

This result can be contrasted to the configuration where the central line is cut [Fig.~\ref{fig_disp_qhe}(c)]. In that case, the M\"obius edge and the central line become equivalent because they have the same shape. No hybridization occurs between the two copies of the central-line edge states, i.e., the blue dispersions remain degenerate as well. In fact, the complete dispersion is a two-fold degenerate copy of the honeycomb ribbon at $1/3$ flux,\cite{BeugelingEA2012PRB86-07} because at each longitudinal coordinate, we have two equal honeycomb ribbons under that same flux. 

It is important to realize that the domain-wall and the cut case share their topological properties. The hopping amplitude at the central line can be tuned adiabatically from $t$ to $0$, without closing any of the bulk gaps. Thus, no topological transition takes place: The number of edge states inside each bulk gap remains the same. The same is true if one compares the domain-wall configuration for the even and odd case. The distinction between these two cases is merely due to the possibility of a cut; in odd-width zigzag ribbons, the cut cannot be made. The similarity of the topological properties is expected with the bulk-boundary correspondence in mind.

\subsection{Quantum spin Hall effect on a M\"obius band with intrinsic spin-orbit coupling}


\begin{figure}[t]
  \includegraphics[width=84mm]{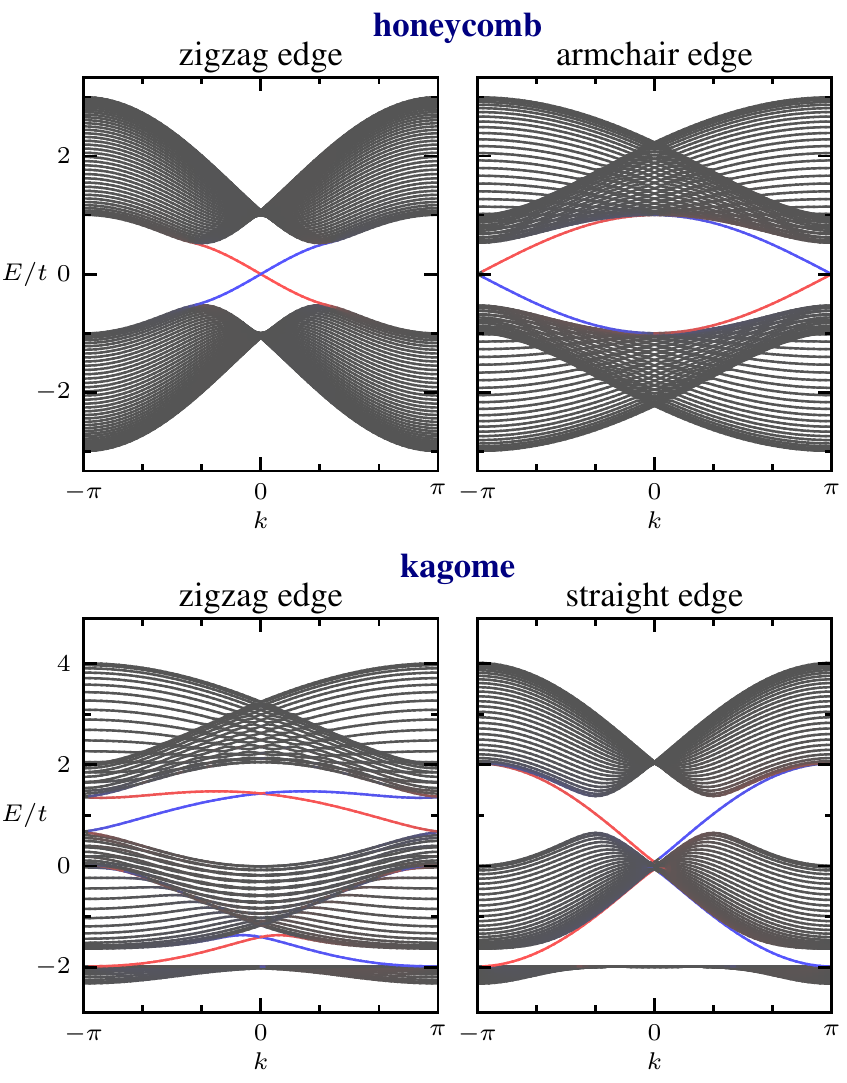}
  \caption{(Color online) Dispersions of the M\"obius band made from a honeycomb lattice with zigzag and armchair edges, and made from a  kagome lattice  with zigzag and straight edges. The colors red and blue indicate the QSH-like edge states, i.e., two opposite ``spins'' propagating in opposite directions on the edge. The strength of the intrinsic SO term is set to $t_\mathrm{I} = 0.1t$.}
  \label{fig_disp_qshe}
\end{figure}


It has already been argued that a QSH-like state exists on a M\"obius strip, both by considering the Haldane model for the graphene case, and more generally, by looking at the topological classification in Ref.~\onlinecite{ChiuEA2013}. Since the intrinsic SO coupling term in the Hamiltonian is reflection symmetric [see \eqn\eqref{eqn_hiso}], its implementation is relatively straightforward; no domain walls of any kind are necessary. Using the methods described in Sec.~\ref{subsec_band_structure_method}, we calculated the dispersions for a M\"obius band with intrinsic SO coupling. In Fig.~\ref{fig_disp_qshe}, the results are shown for the honeycomb and kagome lattice structures; in both cases the two possible edge configurations are shown. The colors in Fig.~\ref{fig_disp_qshe} are determined from the product of the spin and location: The red curves indicate the two right-moving edge currents on the bottom edge of one and on the top edge of the other component of the double cover (which corresponds to the opposite spin). Similarly, the blue curves correspond, to the left-moving edge currents, cf.\ Fig.~\ref{fig_qshe_double_cover}(a). After the $T$-odd states have been projected out, we observe only a single (i.e., nondegenerate) left-moving and a single right-moving edge current. Since there are no hybridization effects, this projection merely removes the two-fold degeneracy of the cylindrical spectrum. These results agree with the evidence for the existence of QSH-like states based on topological arguments of Sec.~\ref{sect_topological_arguments} and based on the extended classification of topological invariants.

Rashba SO coupling in the honeycomb ribbon [\eqn\eqref{eqn_hrso}] can be included in a straightforward manner. As expected from the methods described above, it has a similar effect as in the cylindrical case: Rashba SO coupling by itself does not open a gap. In the presence of intrinsic SO coupling, the Rashba coupling decreases the size of the topological gap or opens a trivial one, depending on its strength.\cite{KaneMele2005PRL95-14} The conservation of vertical spin is broken in this case, but this does not destroy the QSH-like state, similar to what is observed for the cylinder.

\section{Discussions}%
\label{sect_discussion}%
In this paper, we show how topological effects can be described on M\"obius bands by using the ODC of the band. This procedure allows us to relate  the M\"obius system to a system on the cylinder with a reflection symmetry. Through this mapping, a topological classification for states on the M\"obius band was obtained. In general, a QH phase does not exist on the M\"obius, but a QSH state is possible. 

Furthermore, we provide a systematic treatment of the various Hamiltonians that can be fitted seamlessly on a M\"obius band, with examples for different kinds of lattices, such as honeycomb (graphene), Lieb, kagome, and dice lattices. In addition, a physical motivation for the absence or existence of QH and QSH effects in the graphene lattice was given as an example, based on the Hamiltonian symmetries; the magnetic term that usually generates a QH effect cannot exist on the M\"obius strip while the intrinsic SO coupling that generates the QSH effect can. Furthermore, the possibility of inducing chiral edge states without time-reversal symmetry by using magnetic fields with domain walls was explored in detail, to show the interplay between the nontrivial topology of the M\"obius band, and external interactions.

The experimental setup for the observation of the QH effect on a M\"obius band can be envisaged in the following way: Let us consider a graphene layer, and add an insulating layer on top of it. Then, through the longitudinal center of the band [the blue curve in Fig.~\ref{fig_qhe_cut}(a)] we attach a lead, which is isolated from the graphene sample by the insulating layer. The next step is to fold the graphene ribbon by performing  a twist and to bind the ends of the ribbon, such that a M\"obius band is created. Now, we drive a strong current through the lead. As a consequence, a magnetic field will be generated, which is ``entering'' in the lowest part of the band, and ``exiting'' in the upper part. Because the lead has been folded together with the graphene ribbon, the so constructed magnetic field has the properties studied here, and generates a longitudinal domain wall, as shown in Fig.~\ref{fig_qhe_cut}(a). 

Although we have provided a thorough description of topological states of matter in nontrivial surfaces, such as the M\"obius band, there is still much left to future investigations. One possible venue is to describe other nonorientable systems, of a different nature or dimensionality, using the extended topological classification. Many of the possible topological phases on the M\"obius strip, particularly those belonging to topological superconductors, have not been examined yet. Finally, the investigation of topological defects and more intricate types of domain walls, as well as effects of curvature, represent fascinating themes, for which this work will serve as a basis. 

\acknowledgments
We thank V.\ Juri\v{c}i\'c for useful discussions. The work by A.Q.\ is part of the D-ITP consortium, a program of the Netherlands Organisation for Scientific Research (NWO) that is funded by the Dutch Ministry of Education, Culture and Science (OCW). C.M.S.\ acknowledges NWO for funding within the framework of a VICI program.


%

\end{document}